\documentclass[10pt,a4paper,notitlepage]{article}

\usepackage{geometry}  
\usepackage{extsizes}
\usepackage{amssymb,amsmath,amsthm}

\usepackage{multirow}
\usepackage{url}

\usepackage{placeins}
\usepackage{graphicx}
\usepackage{subfigure}

\newcommand\be{$$}
\newcommand\ee{$$}
\newcommand\ben{\begin{equation}}
\newcommand\een{\end{equation}}
\newcommand\bea{\begin{eqnarray*}}
\newcommand\eea{\end{eqnarray*}}
\newcommand\bean{\begin{eqnarray}}
\newcommand\eean{\end{eqnarray}}

\newcommand\kap{{\ensuremath{\rm cap}}}

\newcommand\cdsPrem{{\ensuremath{\rm PremLeg}}}
\newcommand\cdsProt{{\ensuremath{\rm ProtLeg}}}
\newcommand\cdsRelief{{\ensuremath{\rm ReliefLeg}}}
\newcommand\bank{{\ensuremath{\rm bank}}}
\newcommand\Reg{{\ensuremath{\rm Reg}}}
\newcommand\AC{{\ensuremath{\rm AC}}}
\newcommand\rr{{\ensuremath{\rm rec}}}
\newcommand\entity{{\ensuremath{\rm entity}}}
\newcommand\cq{{\ensuremath{\rm c}}}
\newcommand\I[1]{{\ensuremath{I_{\{#1\}}}}}
\newcommand\E{{\ensuremath{\mathbb E}}}

\newcommand\Q{{\ensuremath{\mathbb Q}}}
\newcommand\relief{{\ensuremath{\rm relief}}}

\newcommand\IMM{{\ensuremath{\rm IMM}}}
\newcommand\EAD{{\ensuremath{\rm EAD}}}
\newcommand\PD{{\ensuremath{\rm PD}}}
\newcommand\DCC{{\ensuremath{\rm DCC}}}
\newcommand\CVC{{\ensuremath{\rm CVC}}}
\newcommand\RWA{{\ensuremath{\rm RWA}}}
\newcommand\LGD{{\ensuremath{\rm L_{GD}}}}
\newcommand\cov{{\ensuremath{\rm cov}}}
\newcommand\h{{\ensuremath{\frac{1}{2}}}}
\newcommand\p{para.}

\begin{document}
\author{Chris Kenyon and Andrew Green\footnote{Contact: chris.kenyon@lloydsbanking.com}}
\title{CDS pricing under Basel III:\\
capital relief and default protection
 \footnote{\bf The views expressed are those of the authors only, no other representation should be attributed.}}
\date{22 November 2012, Version 1.00}

\maketitle

\begin{abstract}
Basel III introduces new capital charges for CVA.  These charges, and the Basel 2.5 default capital charge can be mitigated by CDS.  Therefore, to price in the capital relief that CDS contracts provide, we introduce a CDS pricing model with three legs: premium; default protection; and capital relief.  Under simple assumptions we show that 20\%\ to over 50\%\ of observed CDS spread could be due to priced in capital relief.  Given that this is different for IMM and non-IMM banks will we see differential pricing?
\end{abstract}

\section{Introduction}

Basel III \cite{BCBS-189} updates capital charges, and introduces new ones such as CVA (Section VIII \p{97}-\p{99}), which can be mitigated with CDS.  Observers have noted that this linkage can create a so-called doom loop where uncollateralized exposures drive counterparties to buy CDS that then push out CDS spreads, which make unprotected exposure more expensive, etc. \cite{Pollack2011a, Murphy2012a}.  This loop, which is driven by capital costs not changes in default probability, is one reason given for the exemption of sovereign CVA (few post collateral, e.g. Portugal, Ireland) from one CRD IV draft \cite{Cameron2012b}\footnote{At time of writing CRD IV is not final}.  Therefore, to price in the capital relief that CDS contracts provide in addition to default protection, we introduce a new CDS pricing model with three legs: premium; default protection; and capital relief.  Of course capital relief is only present whilst the reference entity has not defaulted.  This also implies that the translation from observed CDS spreads to market-implied default probabilities must include an adjustment for capital relief.  This adjustment can be calculated using our model.

If markets are complete, with no CDS bond basis, then CDSs can be replicated by taking short positions in risky floating bonds issued by the reference entity and a riskless bank account \cite{Carr2005a}.  If these conditions do not hold, then it is theoretically possible that the capital relief that CDSs bring will be priced in.  Thus our model provides bounds on the CDS-implied hazard rates when markets are incomplete.  

No knowledge of the CDS buyer's exposure profile is required by the CDS seller because the buyer's choice of CDS notional and tenor provide a lower bound.  However, the portfolio is important when we start from a particular profile in our examples.  We analyse the standardized CVA charge formula  \cite{BCBS-189}, \p{104}, and show that it is consistent with all counterparties having mutual correlations of 25\%.  This then enables us to compare with a hypothetical IMM case where the mutula correlations are different.

When CDS prices include capital relief, there is no longer a unique CDS price because different banks have different approved regulatory capital calculations (the key results are summarized in Table \ref{t:attrib}).  Whilst a bank's regulatory status is not public, it is generally known to counterparties.  In addition the capital relief will depend on what is being hedged, in the Current Exposure Methodology there are different addons for rates, FX, precious metals, and other commodities.  Of course the asset class being hedged may or may not relate to the main businesss of the CDS reference entity.    We show that CDS buyer regulatory status and the asset class being protected can have significant effect even in simple examples, especially at shorter ($<5$Y) maturities.  

Our treatment of the CVA capital charge under Basel III for IMM banks is limited to the EAD profile (and some commentary on correlation and stressed parameters).  We assume that the 3x factor in VaR (footnote 37 on \p{100}) roughly equates to the difference between the VaR horizon and the one year horizon (\p{104}) used in the standardised CVA risk capital charge.   Will we see differential pricing post Basel III?

\section{CDS under Basel III} 

In our setup a bank buys a CDS on a reference entity from a CDS counterparty.  We make the following assumptions.
\begin{itemize}
\item Capital relief obtained under Basel 2.5 \cite{BCBS-128} (Part 2: II.D.5) for default risk, and under Basel III \cite{BCBS-189} (\p{102}, \p{103}) for CVA risk, for the lifetime of the CDS hedges is priced into CDS spreads, so we have an additional capital relief leg.
\item CDS are traded with perfect collateralization (no minimum transfer amount, no threshold, instantaneous transfers).
\item There is no correlation between the collateralized CDS counterparty default and the CDS reference entity default.
\end{itemize}
The assumptions on CDS collateralization and default correlation are also to avoid the issues pointed out in \cite{Brigo2010b} for counterparty risk on CDSs.

A CDS spread (for simplicity we operate with CDS spreads rather than upfront plus standardized spread levels) is fair when premiums are equal to default protection plus capital relief:
\ben
\cdsPrem_{a,b}(\rr) = \cdsProt_{a,b}(\rr) + \cdsRelief_{a,b}(\Reg(\bank),\entity,\AC)
\label{e:legs}
\een
The capital relief leg depends on both the default reference entity and the buyer's (\bank) regulatory status $\Reg()$ because different buyers will obtain different capital relief depending on their regulatory status.  Note there is no portfolio dependence w.r.t. the buyer's deals with the reference entity because that is implicitly included in the choice of the CDS notional, tenor, etc.

When the bank is not IMM approved it may be calculating \EAD\ using a method where the \EAD\ depends on the asset class (\AC) of the underlying.  For the Current Exposure Methodology (CEM) \cite{BCBS-128} (section VII, \p{91}-\p{96}) Table \ref{t:cemWts} shows the dependence on the underlying asset class.

\begin{table}[tb]
	\centering
		\begin{tabular}{c|ccccc}
		Maturity & IR & FX/Gold & Equities & PM & OC \\ \hline
		$\le$1Y   & 0.0\% & 1.0\% & 6\% & 7\% & 10\% \\
1Y$<$ and $\le$5Y & 0.5\% & 5.0\% & 8\% & 7\% & 12\% \\
		$>$5Y	& 1.5\% & 7.5\% & 10\% & 8\% & 15\% 
		\end{tabular}
	\caption{CEM potential future exposure notional add-on dependency on maturity and asset class under \p{92} of \cite{BCBS-128}.  PM = precious metals other than Gold. OC = other commodities.}
	\label{t:cemWts}
\end{table}

\begin{figure}[htbp]
	\centering
		\includegraphics[width=0.7\linewidth]{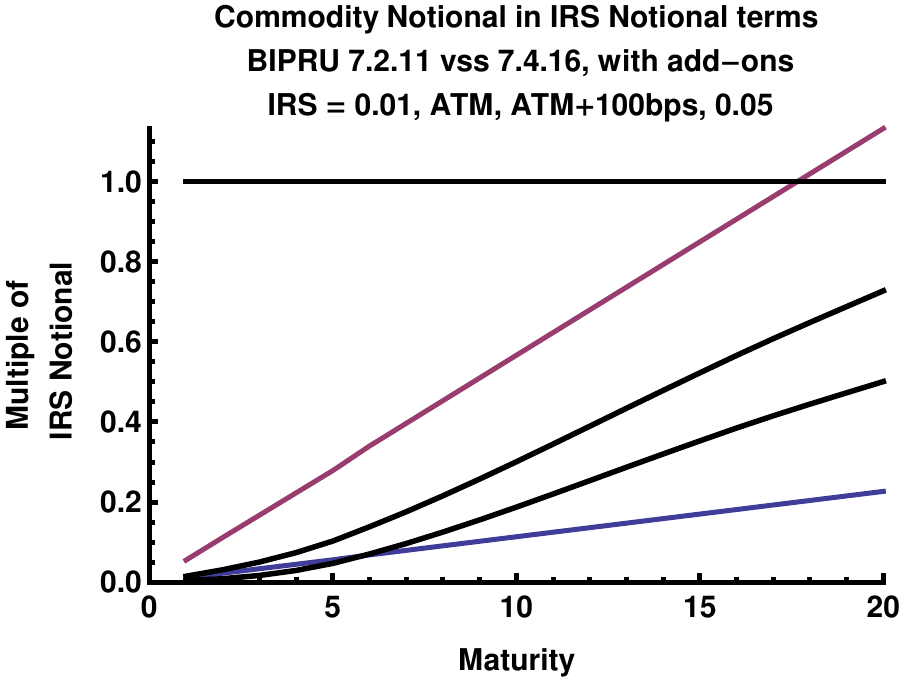}
	\caption{Commodity notional interpreted as IRS notionals (including CEM addons) for a range of IRS levels. For current IRS levels (curved black lines are ATM and ATM+100bps) commodities have relatively low regulatory notionals.}
	\label{f:oilASirs}
\end{figure}

Things are not quite as Table \ref{t:cemWts} may suggest because different asset classes have different definitions of the notional of the trade.  The notional of an interest rate swap is the quantity used to create coupon payments (BIPRU 7.2.11).  However, the notional of an other commodity (OC) trade, e.g. WTI Oil, is the sum of the commodity flows (BIPRU 7.4.16).  Hence OC notionals are much smaller than IR notionals for medium to longish maturities (in IRS terms).  Figure \ref{f:oilASirs} shows the relative advantage of OCs vs interest rate levels for maturities out 20 years.  Of course the regulatory notional of an IRS does not depend on its tenor (coupon frequency) which is also an approximation.

Equation \ref{e:legs} uses hazard rates on all three legs because capital relief is only valuable whilst the reference entity has not defaulted.  For an IMM bank the CDS rate is used in both the premium leg and the relief leg because CVA VaR uses observed CDS spreads (not capital-adjusted CDS spreads).  Thus IMM banks are, for this item, at a relative disadvantage to non-IM banks whose CVA formula (Equation \ref{e:standcva}) does not use observed CDS spreads.

We now expand each leg in Equation \ref{e:legs}.
\bean
\cdsPrem_{a,b}(\cq) &=& \cq\ \E[D(0,\xi) (\xi - T_{\beta(\xi)-1})  \I{T_a<\xi<T_b}  ]  \nonumber \\
&& {}+ \sum_{i=a+1}^b \cq\ \E[ D(0,T_i)\tau_i \I{\xi\ge T_i}  ] \nonumber \\
&=& \cq \int_{T_a}^{T_b} P(0,t) (t - T_{\beta(t)-1}) \Q(\xi\in[t,t+dt])\nonumber\\
&& {}+ \cq \sum_{i=a+1}^b P(0,T_i)\tau_i \Q(\xi\ge T_i)\label{e:prem} \\
\cdsProt_{a,b}(\LGD) &=& \E[\I{T_a <\xi \le T_b} D(0,\xi)\LGD ]\nonumber  \\
&=& \LGD \int_{T_a}^{T_b} P(0,t) \Q(\xi\in[t,t+dt]) \label{e:prot} \\
\cdsRelief_{a,b}(K(.))&=&  \int_{T_a}^{T_b}\E[ D_\kap(0,t) H(K_\relief(t,\cq\I{\IMM}),t)\I{\xi\ge t}  ] \nonumber \\
&=& \int_{T_a}^{T_b} P_\kap(0,t) H(K_\relief(t,\cq\I{\IMM}),t) \Q(\xi\ge t) \label{e:relief}
\eean
Where:
\begin{itemize}
\item $a,b$ protection limit times for the CDS;
\item $\beta(\xi)$ number of next coupon payment after time $\xi$;
\item $\cq$ CDS spread (rate);
\item $D(0,t)$ stochastic riskless discount factor from $0$ to $t$;
\item $D_\kap(0,t)$ stochastic capital discount factor from $0$ to $t$;
\item $P(0,t)$ riskless zero coupon bond with maturity $t$;
\item $P_\kap(0,t)$ capital zero coupon bond with maturity $t$;
\item $K_\relief(t,\cq\I{\IMM})$ capital relief from unit notional of CDS protection at time $t$; this depends on the {\it observed} CDS spread for IMM banks;
\item $H(.,t)$ instantaneous cost of capital at $t$;
\item $\tau_i$ year fraction for $i$th premium payment;
\item $\Q(.)$ survival probabilities at time $0$.
\end{itemize}
Equations \ref{e:prem} and \ref{e:prot} are standard \cite{Brigo2006a} under the assumptions given above, Equation \ref{e:relief} is new to capture the capital relief obtained from CDS contracts.  Depending on the circumstances it is possible that not all the capital relief is priced in, this gives a maximum relief value for DCC and CVC.

  We have assumed zero transaction costs, e.g. for changing levels of capital.  The equivalence in Equation \ref{e:relief} is when the bank has a given cost of capital, or target rate of return on capital which can then be independent of interest rates and reference entity default time.

The fair CDS spread from Equations \ref{e:legs}, \ref{e:prem}, \ref{e:prot}, \ref{e:relief} is:
\ben
c=\frac{\LGD \int_{T_a}^{T_b} P(0,t) \Q(\xi\in[t,t+dt]) + \int_{T_a}^{T_b} P_\kap(0,t) H(K_\relief(t,\cq\I{\IMM}),t) \Q(\xi\ge t)}{\int_{T_a}^{T_b} P(0,t) (t - T_{\beta(t)-1}) \Q(\xi\in[t,t+dt])
+ \sum_{i=a+1}^b P(0,T_i)\tau_i \Q(\xi\ge T_i)
}\quad \label{e:fair}
\een
For CDS buyers with IMM approval Equation \ref{e:fair} has the fair CDS spread appearing on {\it both} sides of the equation as it is used (by regulation) in the CVA capital charge.  Thus in that case Equation \ref{e:fair} requires non-linear numerical solution. 

\section{Capital Pricing}

For simplicity we start  from the point of view of a non-IMM bank and consider only credit risk capital, i.e. default capital (leading to a default capital cost, DCC) and CVA VaR capital (leading to a CVA capital cost CVC). We do not include market risk or operation risk, etc.  Where there are ambiguities in the Basel documents we use UK regulations (BIPRU) for details.  We go into depth on the derivation of the regulatory equations for the CVA capital charge to understand the (effectively frozen) portfolio effects in a non-IMM bank, and how an IMM bank's (variable) portfolio characteristics can result in different capital charges.

Basel III specifies the capital required at any given date.  However, the cost of capital for a trade is the lifetime capital cost, not the cost of the trade-date capital requirement.  We consider all capital costs in terms of lifetime cost.  Of course this lifetime depends on the lifetimes of the counterparties.  We take the point of view that the bank (or trader) considers costs as a going concern so uses counterparty default time as the end of the trade if this occurs prior to maturity.  It would be possible to include own-default time and this is a straightforward extension.

\subsection{CVA Capital Charge}

We start from the standardized CVA risk capital charge in \cite{BCBS-189}, \p{104}, noting that this is not RWA but capital directly:
\bean
K_\CVC &=&2.33\sqrt{h}\left\{ \left(\sum_i 0.5 w_i\left(M_i\EAD_i-M_i^{\rm hedge}B_i\right) - \sum_{\rm ind} w_{\rm ind}M_{\rm ind}B_{\rm ind}\right)^2 \right. \nonumber \\
&&
\left.\qquad\qquad\vphantom{\sum_{\rm ind}}   {}+ \sum_i 0.75 w_i^2\left(M_i\EAD_i^{\rm total}-M_i^{\rm hedge}B_i\right)^2  \right\}^{1/2} \label{e:standcva}
\eean
Where:
\begin{itemize}
\item $h$ one year risk horizon in units of years, i.e. h=1;
\item $w_i$ risk weight of $i^{\rm th}$ counterparty based on external rating (or equivalent);
\item $\EAD_i^{\rm total}$ exposure at default of counterparty $i$, discounted using $\frac{1-e^{-0.05 M_i}}{0.05 M_i}$ (as we are using the non-IMM point of view);
\item $B_i$ notional of purchased single name CDS hedges, discounted as above;
\item $B_{\rm ind}$ notional of purchased index CDS hedges, discounted as above;
\item $w_{\rm ind}$ risk weight of index hedge using one of seven weights using the average index spread;
\item $M_i$ effective maturity of transactions with counterparty $i$, for non-IMM this is notional weighted average, and is not capped at five years;
\item $M_i^{\rm hedge}$ maturity of hedge instrument with notional $B_i$;
\item $M_{\rm ind}$ maturity of index hedge ${\rm ind}$.
\end{itemize}
It is clear that Equation \ref{e:standcva} is derived from two sources: firstly a 99\%\ 1-sided Standard Normal distribution with mean zero, gives the 2.33 factor; secondly there is an assumption that all counterparties have a correlation of 25\%.  Taking Equation \ref{e:standcva} with no hedging we have:
\bea
K^2 &\propto& \left(\sum_i 0.5 w_i M_i\EAD_i\right)^2 + \sum_i 0.75 w_i^2 M_i^2\EAD_i^2 \\
&=& \left(\h\sum_i \sigma_i\right)^2 + \frac{3}{4} \sum_i \sigma_i^2 \\
&=& \frac{1}{4} n^2\sigma^2 +  \frac{3}{4} n \sigma_i^2 \\
&=& \sigma^2\left(\frac{1}{4} n^2 +  \frac{3}{4} n \right) 
\eea
Where we have written $\sigma_i=w_i M_i\EAD_i$ and then made the assumption that all the $\sigma_i$ are equal.  Now consider the variance $V(n,\rho)$ of $n$ random variables with mutual correlation $\rho$
\bea
V(n,\rho) &=& \sum_{i=1}^n \sum_{j=1}^n \cov(i,j) \\
&=& \sum_{i=1}^n \sigma_i^2 + 2\sum_{i=1}^n \sum_{j=i+1}^n\rho_{i,j}\sigma_i\sigma_j \\
&=& n \sigma^2 + n(n-1)\rho \sigma^2 \\
&=& \sigma^2\left(\rho n^2 + n(1-\rho)\right)
\eea
Hence $\rho=1/4$, after making similar assumptions on $\sigma_i$ for the $n$ random variables.  As $n$ increases the proportionality factor for $K$ quickly converges on $1/2=\sqrt{1/4}$ as $n^2$ soon dominates $n$.

We note that equal mutual {\it positive} correlations always lead to a valid correlation matrix whatever the number of counterparties (there are $n-1$ eigenvalues $1-\rho$ and one eigenvalue $(n-1)\rho$), so $\rho=0.25$ is valid.  

We can now ask how CVA VaR capital depends on the portfolio distribution.

\subsubsection{Portfolio Effects}

\begin{figure}[htbp]
	\centering
		\includegraphics[width=0.6\linewidth]{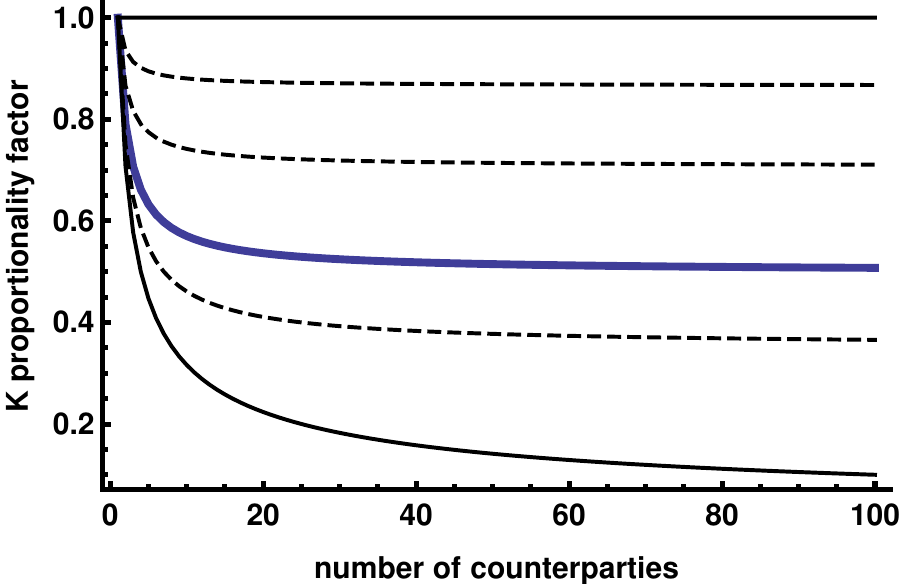}
	\caption{Equal counterparty contributions.  Relative proportionality factor $V(n,\rho)$ from Equation \ref{e:standcva}, compared to $\rho=0.25$ (regulatory value) for $\rho=0.00$ (full line) $0.1$ (dashed) $0.25$ (thick line) $0.50, 0.75$ (dashed) and $1.00$ (full line).  }
	\label{f:sCVAvar}
\end{figure}

Figure \ref{f:sCVAvar} shows the how the proportionality factor for K in Equation \ref{e:standcva}, $V(n,\rho)$, depends on $n$ and $\rho$.  Of course Equation \ref{e:standcva} has a fixed value of $\rho$ built in.  Portfolios with higher $\rho$ will be charged lower capital with the standardized formula than if their actual CVA correlation was used. Figure \ref{f:sCVAvar} shows this when counterparties are all equal size.  Note that the regulatory value, $\rho=0.25$ means that each counterparty's capital effect is close to $1/2$ of its stand-alone effect.

Usually counterparties are not of equal size but follow a distribution.  We assume that this portfolio distribution is Log-Normal (limiting distribution when underlying factors are multiplicative).  We keep the average size (volatility or $\sum w_i M_i\EAD_i$) constant and alter the dispersion parameter $\sigma_D$ of the portfolio distribution $D$, thus
\be
D\sim e^{\mu_D-\sigma_D^2/2+\sigma_D N}
\ee
Where $N$ is a standard Normal distribution, and $\exp(\mu_D)=1$ (arbitrary choice).
Individual counterparty sizes are taken as quantiles of $D$.  The number of quantiles used is given by the number of counterparties.  Since the counterparties are now not of equal size the exact equivalence with $\rho=0.25$ no longer holds.  

\begin{figure}[htbp]  
\begin{minipage}[b]{0.45\linewidth}  
\centering 
\includegraphics[width=0.9\textwidth]{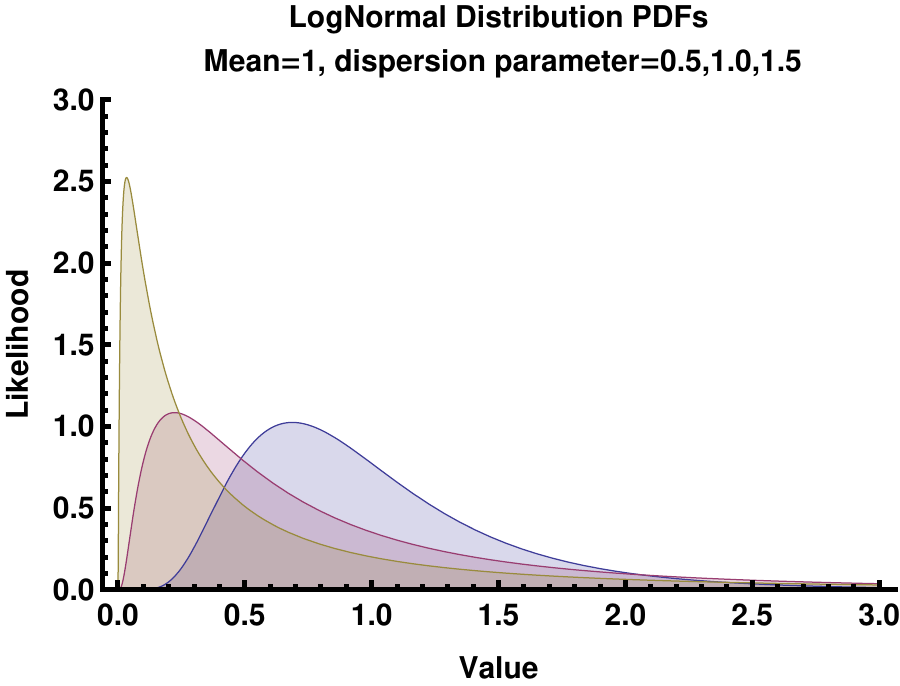}  
\end{minipage}\hspace{0.0cm}\begin{minipage}[b]{0.45\linewidth}  
\centering 
\includegraphics[width=\textwidth]{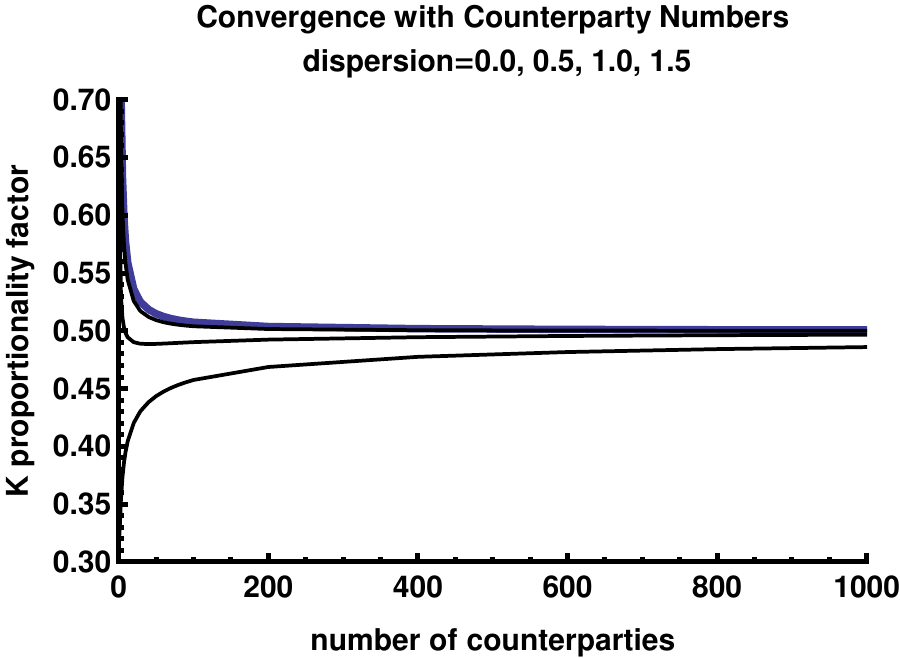}  
\end{minipage} 
\caption{For a range of LogNormal counterparty $w_iM_i\EAD_i$ distributions ({\bf left panel}, dispersion increases as peak moves towards the origin) the proportionality factor in Standardized CVA capital charge converges towards 1/2 as $n$ increases ({\bf right panel}).  The thick line corresponds to $\sigma_D=0.0$ and obscures the line with $\sigma_D=0.5$, the lowest line has $\sigma_D=1.5$.} 
\label{f:sCVAvarConverg} 
\end{figure} 

Figure \ref{f:sCVAvarConverg} shows how the proportionality factor in the standardized CVA capital charge converges as the number of counterparties $n$ increases.  With a range of dispersions of the counterparty sizes, $\sigma_D=0.5,\ 1.0,\ 1.5$ we see that the proportionality factor converges to around 1/2 for reasonable numbers of counterparties, i.e. around 1000.  The dispersion parameter of  $\sigma_D=1.5$ gives a long tail of counterparty sizes, and the other cases model more focused portfolios of counterparty sizes.

\begin{figure}[htbp]
	\centering
		\includegraphics[width=0.6\linewidth]{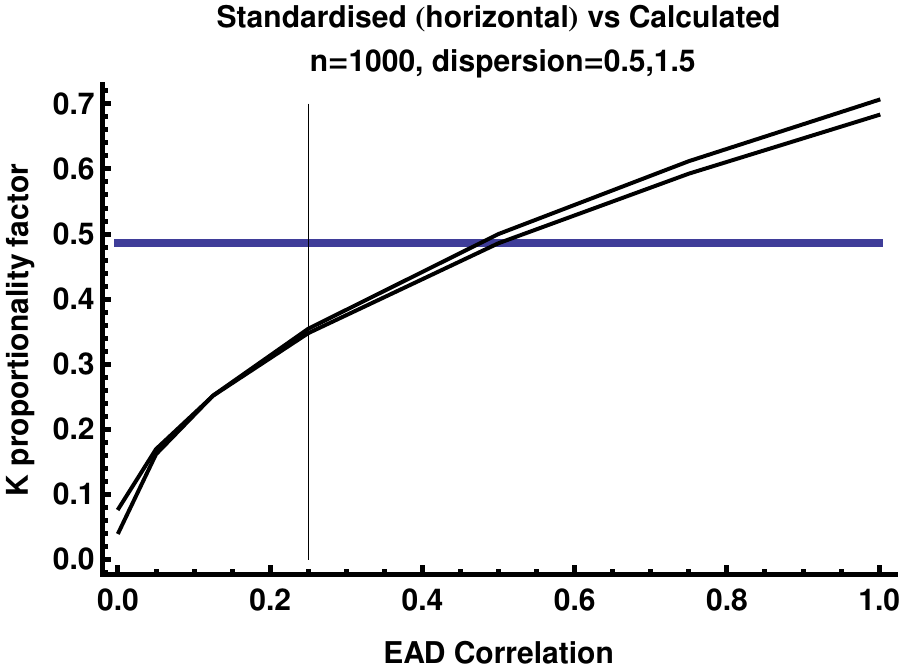}
	\caption{dispersed counterparty contributions.  Relative proportionality factor of standardized CVA (horizontal line, corresponds to 25\%\ correlation, vertical line) compared with calculation for correlated counterparties.  Standardized CVA is conservative relative to counterparties with a mutual correlation of up to 50\%.}
	\label{f:sCVAvsCalc}
\end{figure}

Figure \ref{f:sCVAvsCalc} considers the standardized CVA proportionality factor for $n=1000$ and compares with a set of counterparties with differing (common) correlations.  We see that the standardized calculation is conservative for correlations up to about 50\%\ and not conservative thereafter.  Now in a crisis correlations can increase substantially. This range of $K$ between benign and high-correlation scenarios is captured by the fact that IMM banks must use the {\it sum} of stressed and non-stressed capital charges.  Thus non-IMM banks may actually have the advantage.  Here we assume for simplicity that the sum of the stressed and non-stressed parameters effects is the same as the non-IMM $K$ factor as it is a simple multiplier.

\FloatBarrier
\subsection{Default Capital Charge}

This is mostly unchanged from Basel 2.5 \cite{BCBS-128} \p{272} and \p{38} of Annex 4, we reproduce the equations in the Appendix for convenience, keeping to their notation.  For the Default Capital Charge (DCC) $M$ (effective maturity) is capped at five years, whereas for CVA Capital Charge $M$ is not capped.  For exposure calculations an IMM bank multiples the EAD of a netting set by a constant factor $\alpha$, whereas a non-IMM bank using CEM uses addons based on notionals.

\p{285} specifies that default probability, \PD, is the one-year probability of default (not CDS spread) and floored at 3bps.  The derivation of \PD\ is given in \p{461}--\p{463}.  Three techniques are permitted (internal default experience, mapping to external data, and statistical default models) and banks are required to take appropriate account of long-run experience, explicitly stated to be at least five years \p{463}.  By implication the major consideration for \PD\ involves historical data not market CDS contracts (unlike CVA capital in \p{98} of Basel III for IMM-approved banks).

\FloatBarrier
\section{CDS Examples}

In some examples we now consider the magnitude of the capital relief leg, and hence the proportion of CDS spread that is paying for capital relief not default protection.  We use the relation that CVA on a trade is equal to the CCDS cost where the CCDS only considers default.  Whilst we acknowledge that in general CVA on swaps is recursive \cite{Burgard2011a} we consider the non-recursive version as an approximation to focus on capital relief.

To get the dependence on market observed CDS spread into the calculations that are based on the standardised CVA capital charge formula we scale the weight $w$ by the ratio of the observed-implied to calculated-implied default probabilities, at $M$.  

\subsection{Interest Rate Swap\label{s:IRS}}

As a basic financial instrument we look at vanilla Interest Rate Swaps (IRS).  For simplicity we assume that counterparty default is independent of interest rates.  The key thing we need to calculate is EAD as this feeds into all the capital calculations, as well as the usual CVA.

\begin{figure}[htbp]
	\centering
		\includegraphics[width=0.6\linewidth]{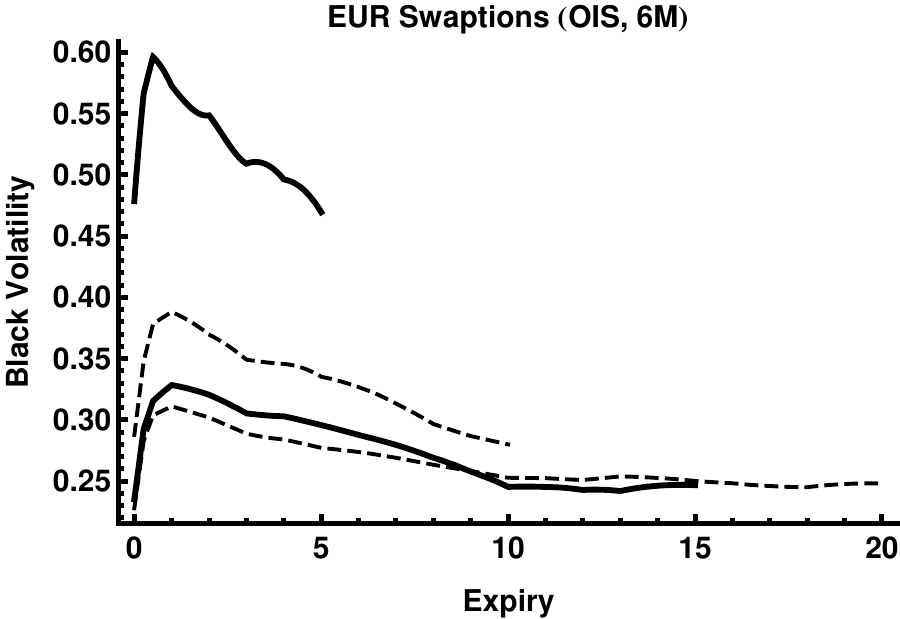}
	\caption{EUR swaption volatilities (6M tenor, OIS discounting) for ATM strikes of swaps with maturities of 5Y, 10Y, 15Y, 20Y (some lines (10Y, 20Y) are dashed for clarity).  Swaption smile is included in the volatilities.}
	\label{f:vol}
\end{figure}

Since default is independent of interest rates the expected exposure at any future time $S$ discounted to the present is given by the corresponding swaption price.  We use the inverse riskless discount factor to $S$ to get the forward premium which is the expected future exposure.  Practically we could obtain forward premia directly from the market, but for examples we use a swaption implied volatility surface (all data from Bloomberg).  Figure \ref{f:vol} shows volatilities holding expiry+maturity constant, that is, corresponding to different underlying swaps.

\begin{figure}[htbp]
	\centering
		\includegraphics[width=0.6\linewidth]{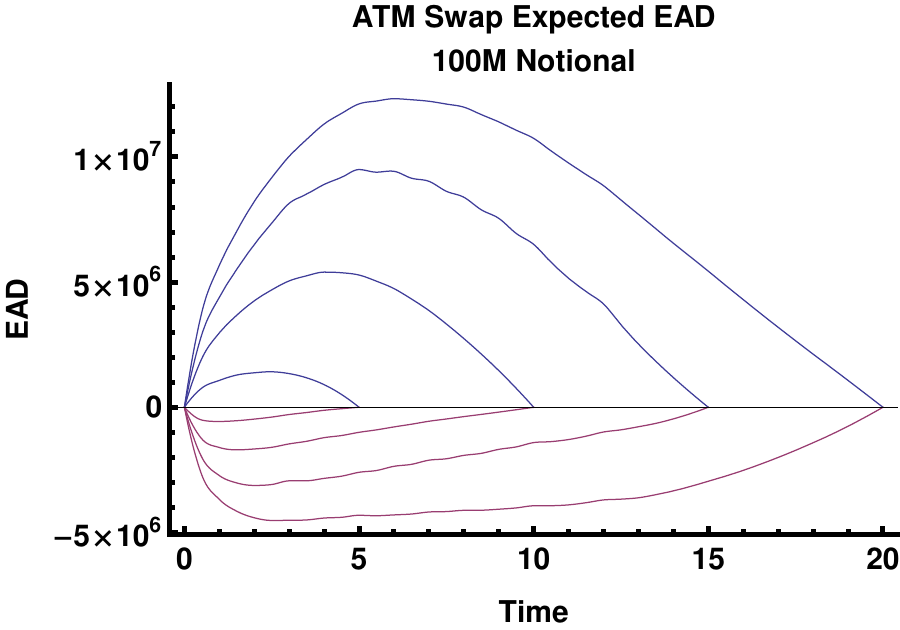}
	\caption{ATM EUR swap risk-neutral expected EAD for 100M notional, with no addons, i.e. without any additions for capital purposes. Exposures have not been discounted back to time zero, and exclude coupon accruals.  Positive curves are for 5Y, 10Y, 15Y, 20Y swaps receive float.  Negative curves are for receive fixed (plotted as negatives for clarity).}
	\label{f:ead}
\end{figure}

The swaption volatilities in Figure \ref{f:vol} lead to the risk-neutral expected EAD profiles shown in Figure \ref{f:ead}.  At this point we remind readers that that risk factors dynamics behind EAD profiles for DCC must pass historical backtesting \cite{BCBS-185}.  The underlying risk factors are explicitly permitted to be calibrated to either market-implied or historical data.  For a detailed discussion see Chapter 11 of \cite{Kenyon2012a}.

\begin{figure}[htbp]
	\centering
		\includegraphics[width=0.6\linewidth]{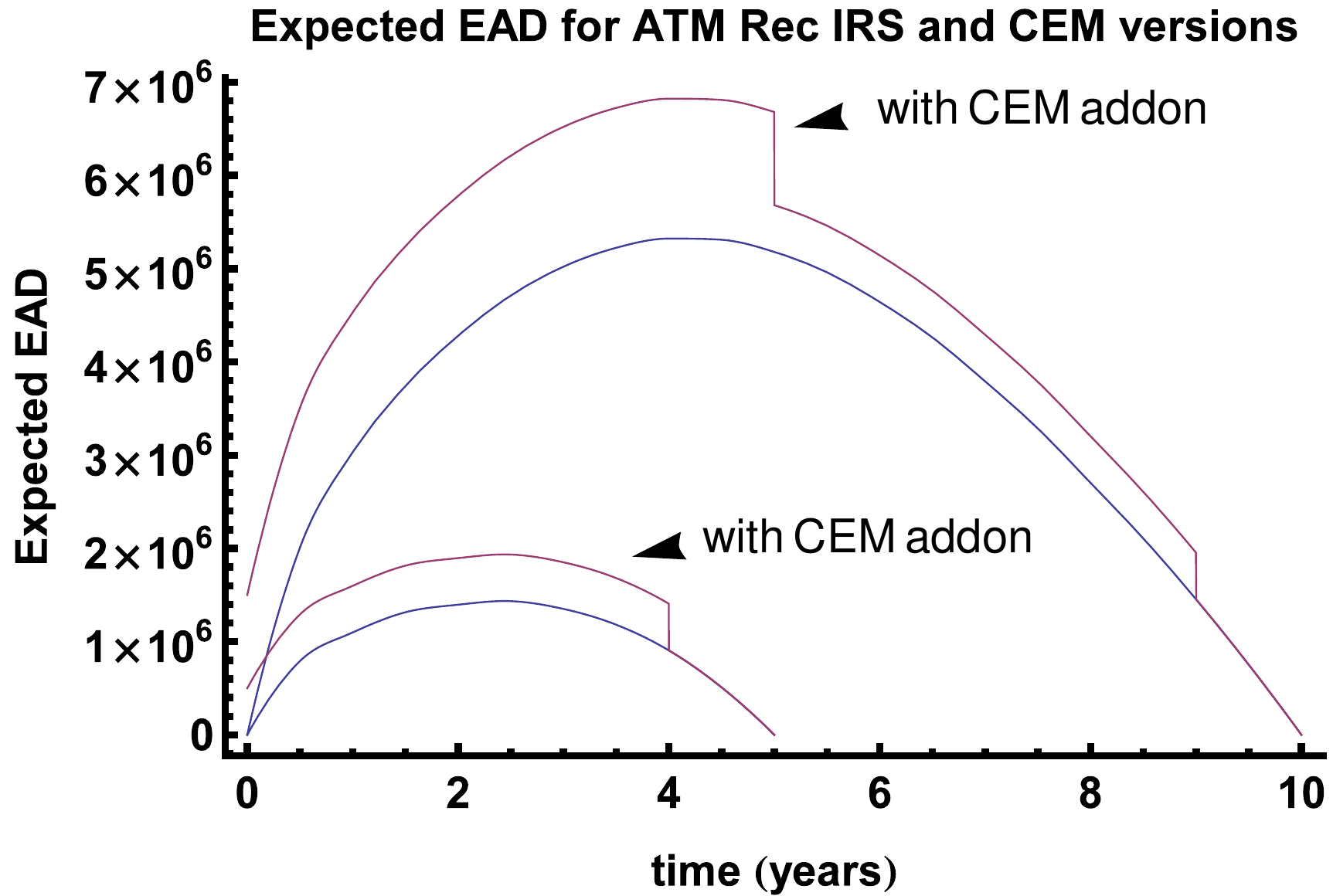}
	\caption{ATM EUR swap risk-neutral expected EAD for 100M notional, with and without CEM addon.  Note that IMM profiles would simply multiply by $\alpha$ (see text for details).}
	\label{f:eadCEM}
\end{figure}

Figure \ref{f:eadCEM} shows the expected EAD profiles for receiver IRS with and without CEM addons that are linked to notionals and to remaining maturities.  With current low interest rates the addons are significant fractions of the profiles.

\begin{figure}[htbp]
	\centering
		\includegraphics[width=0.6\linewidth]{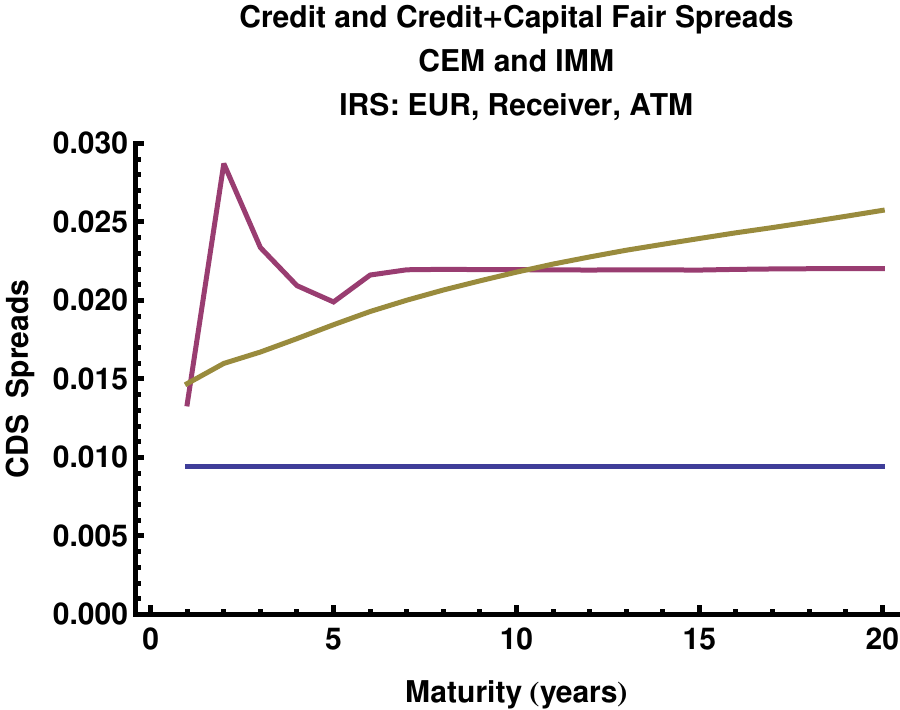}
	\caption{Fair CDS spreads for default-only (horizontal line), CEM capital calculation (jagged line) and IMM capital calculation (increasing line).}
	\label{f:cds}
\end{figure}

\begin{table}
	\centering
		\begin{tabular}{ccc}\hline
			Parameter & Value & Source/Motivation \\ \hline
			Alpha $\alpha$ & 1.3 & middle of range \\
			Hazard rate & 0.0156 & so 5Y observed CEM CDS is 0.02  \\
			Recovery Rate & 0.40 & typical \\
			Historical Default Probability & 0.0024 & global BBB from S\&P\cite{Vazza2012a}\\
			Cost of Capital & 0.10 & choice \\
			Minimum Capital & 0.10 & typical \\
			Capital Discounting & Cost of Capital & choice \\ \hline
		\end{tabular}
	\caption{Parameters for IRS Examples.  The Hazard rate makes the 5Y observed CDS rate 2\%\ assuming capital is priced in and this is calculated by CEM.  The 2\%\ is chosen to roughly line up with Markit BBB 5Y generic CDS spreads.}
	\label{t:par}
\end{table}

Figure \ref{f:cds} shows the fair CDS spreads for default-only, CEM capital calculation, and IMM capital calculation.  These CDS spreads are for underlying EUR ATM IRS as in the previous examples.  The jaggedness of the CEM calculation derives directly from the changes in addon with increasing swap maturity (one year and below there is no addon).  It is also a function of the current low interest rate regime, so the addon appears large.  Table \ref{t:par} provides the parameters for the example.

\subsection{IRS as Other Commodity}

\begin{figure}[htbp]
	\centering
		\includegraphics[width=0.6\linewidth]{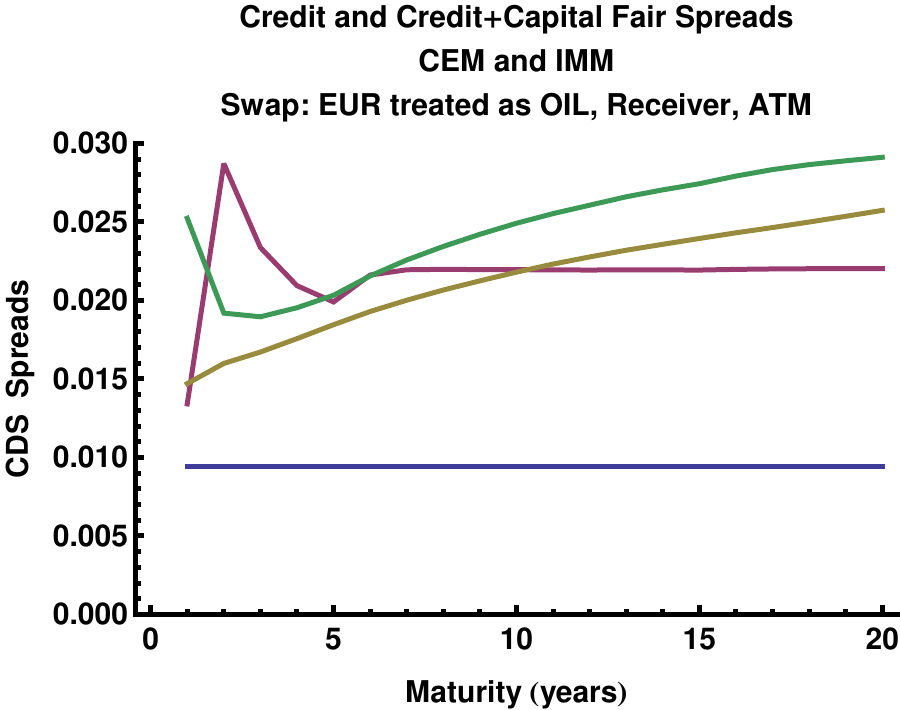}
	\caption{Fair CDS spreads as Figure \ref{f:cds} with an additional line that treats the IRS as though it was an Other Commodity (e.g. Oil) which affects notional calculation for CEM and addons.}
	\label{f:cdsOIL}
\end{figure}

If we now treat the IRS as though it was an Other Commodity we can get an idea of the relative significance of the different notional calculation method and addons.  This is shown in Figure \ref{f:cdsOIL}.  The proportion of CDS spread attributable to capital clearly depends on what asset class is being protected against for non-IMM banks.  Whilst this may appear surprising from a default protection point of view, seen from a capital angle this is a necessary consequence of Table \ref{t:cemWts}.

\subsection{Comparisons}

\begin{table}
	\centering
\begin{tabular}{ccccc|ccc|ccc} \\
\multicolumn{5}{c|}{Parameters} & \multicolumn{3}{c|}{CEM} & \multicolumn{3}{c}{IMM} \\
Rating & CDS & rec & S\&P & $w_i$ & default  & DCC & CVC & default  & DCC & CVC \\ 
 & bps & \% & bps & \% & \% & \% & \% & \% & \% & \% \\ 
\hline
 \text{A} & 90. & 38. & 8. & 0.8 & 27 & 42 & 31 & 38 & 36 & 26 \\
 \text{BBB} & 130. & 38. & 24. & 1. & 18 & 55 & 27 & 29 & 48 & 23 \\
 \text{BB} & 290. & 37. & 90. & 2. & 29 & 47 & 25 & 38 & 42 & 20 \\
 \text{B} & 510. & 36. & 448. & 3. & 34 & 45 & 21 & 41 & 41 & 18 \\
 \text{CCC} & 1170. & 33. & 2600. & 10. & 33 & 36 & 32 & 37 & 35 & 28 \\
\end{tabular}
	\caption{Breakdown of observed CDS spreads for 5Y IRS into default protection, and capital relief: DCC and CVC.  Observed CDS spreads are generic from Markit.  5Y chosen as liquid CDS point.  (OC and Rates capital relief are very close at this maturity (only) so only Rates results shown).}
	\label{t:attrib}
\end{table}

Now we consider reference entities with different hazard rates.  The first case is where the ratings are different (based on S\&P and Markit data) and the second where we hold everything except the hazard rate constant.

Table \ref{t:attrib} shows attribution of observed CDS spreads into default protection, default capital (Basel 2.5) and CVA capital (Basel III), for 5Y IRS.  Both non-IMM and IMM banks are shown.  We do not display the numbers for IRS-as-oil for the non-IMM case because the 5Y maturity is where the asset class is not important (from Figure \ref{f:cdsOIL}).  We pick the 5Y maturity to display because it is usualyy the most liquid CDS spread.  The part of the CDS spread attributable to default protection is less than half, and this holds across a range of ratings (or equivalently observed CDS srpeads).  This proportion is lower than in earlier examples because we have there are higher weightings and the S\&P  long-term default probability increases quickly as rating decreases.

\begin{figure}[htbp]
	\centering
		\includegraphics[width=0.99\linewidth]{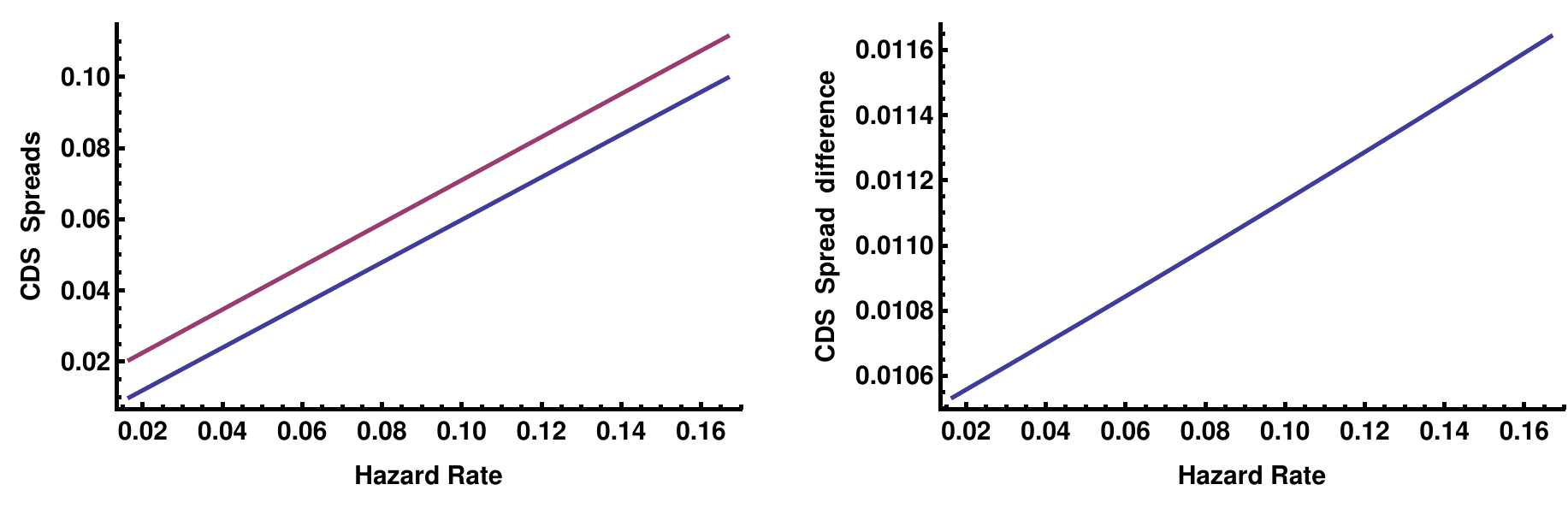}
	\caption{CDS spreads from default protection only and including capital relief.  Subject is EUR 5Y ATM receiver swap with parameters as in Table \ref{t:par}.}
	\label{f:onlyH}
\end{figure}

Figure \ref{f:onlyH} considers CDS spreads when we change {\it only} the hazard rate.  Here we see that the absolute amount of spread attributable to capital relief is nearly constant as the hazard rate increases.   This is unlike Table \ref{t:attrib} where the reference entity changes rating as the hazard rates increase which brings on a set of associated parameter changes.  These parameter changes mean that the capital relief proporation is increasing in absolute terms.

\section{Conclusions}

Under Basel III, and previously under Basel 2.5, CDS provide capital relief.  If capital relief is priced in to CDS prices then a new model is required to price CDSs and derive default probabilities.  We have presented a CDS model that addresses these requirements, now with three legs: premium; protection; and capital relief.  We do not know how much capital relief is actually priced in.  This will be determined by market expectations of when regulations will come into force, market completeness (replication costs), and competition between CDS sellers.  Unless there is a deep liquid market for shorting the reference entity bonds (i.e. practical replication) capital relief pricing is possible.  
  
We have shown that capital relief pricing has a potentially significant effect on CDS spreads, easily being 20\%\ to 50\%\ of the observed CDS spread.  We also showed that both the IMM status of the CDS buyer and the asset class that the CDS buyer is obtaining capital relief on have major effects, especially for shorter maturities (below five years).  In addition institutions on the Systematically Important Banks list \cite{FSB2012b} will see different prices because they have higher minimum capital requirements.  

The capital fraction in CDS prices is highest for non-IMM banks where the counterparty has moderate CDS spreads and a BBB rating.  However, the capital fraction will increase for any rating as the hazard rate decreases.  Low hazard rates may correspond to many safe (low default risk) sovereigns and is consistent with the suggested doom loop for sovereign CDSs \cite{Pollack2011a,Murphy2012a} being driven by capital.

Unlike our non-IMM bank calculations, our IMM-approved bank calculations are approximate in many ways and should be taken cautiously.  Including the observed CDS spread into the standardised CVA calculation via default probability ratio is an approximation.  However, this did show that at the 5Y maturity the effect was at most about a 15\%\ decrease in the default proportion for the CCC case from 42\%\ to 38\%, (data not shown in table).  This effect will increase with maturity.

We have not explicitly made allowance for IMM-banks having to do two calculations: with stressed and non-stressed parameters \cite{BCBS-189}, \p{100}.  By implication we have taken the position that the sum of the stressed and non-stressed $K$ factor effects is equal to the effect of the non-IMM $K$ factor.  There is also the question of VaR horizon and multipliers versus the one year horizon used in the standardized CVA capital calculation.  In general detailed IMM analysis is an area for future investigation. 

For simplicity we assumed that counterparty default is independent of interest rates.  Moving to a fully correlated dynamic model is straightforward using simulation.

Our new CDS model including capital relief can be used to obtain bounds on hazard rates and adjust observed CDS spreads for capital relief.  Given the potential ambiguity of CDS interpretation between default and capital their direct regulatory use for CVA may need reassessment.

\section*{Appendix: Regulatory Formulae}

We reproduce the relevant regulatory equations here for convenience, using the regulatory notation.

\bea
b(\PD)&=&(0.11852 - 0.05478\log(\PD))^2 \\
R(\PD)&=&\frac{0.12(1-e^{50\PD})}{1-e^{-50}} + \frac{0.24(1-(1-e^{50\PD}))}{1-e^{-50}}\\
K_\DCC(\LGD,\PD,M)&=&\left(\LGD\Phi\left(\frac{1}{\sqrt{1-R(\PD)}}\Phi^{-1}(\PD) \right.\right. \\
&&\left.\left.\qquad\qquad{}+\sqrt{\frac{R(\PD)}{1-R(\PD)}} \Phi^{-1}(0.999)  \right)
-\PD\times\LGD
    \right)\\
&&\times\left( \frac{1+(M-2.5)b(\PD)}{1-1.5b(\PD)}   \right)\\
\RWA_\DCC&=&12.5\times K_\DCC\times\EAD  \\
\eea
\be
M = \frac{ \sum_{k=1}^{t_k\le1{\rm year}} {\rm Effective EE}_k \times \Delta t_k \ df_k 
+ \sum_{t_k > 1 {\rm year}}^{{\rm maturity}} {\rm EE}_k \times \Delta t_k \times df_k     }{
\sum_{k=1}^{t_k\le1{\rm year}} {\rm Effective EE}_k \times \Delta t_k \ df_k
}
\ee
The formula for $M$ above is used when the longest contract in the netting set has a maturity greater than one year.  EE is the expected exposure (exposure is by definition floored at zero), and Effective EE uses the maximum EE to date for the first year.  For an internal model on a given date:
\be
\EAD_\IMM = \alpha \times {\rm Effective EPE}
\ee
where EPE is expected exposure on a given date.  $\alpha$ depends on the bank and is generally  limited to the range $1.2$ to $1.4$ (higher is possible but not lower \p{33},\p{34}).  Unlike the CEM addon which adds a multiple of notional, alpha multiplies exposure.

\bibliographystyle{alpha}
\bibliography{kenyon_general}

\end{document}